\documentstyle[prl,aps,graphicx,floats]{revtex}
\begin{document}

\draft

\title{Zeeman and Orbital Limiting Fields:
Separated Spin and Charge Degrees of Freedom in Cuprate
Superconductors}

\author{L.\ Krusin-Elbaum~$^{1,*}$, G.\ Blatter~$^{2}$, and T.\ Shibauchi~$^{3}$}

\address{$^{1}$IBM T.J.\ Watson Research Center, Yorktown Heights, New York 10598, USA}
\address{$^{2}$Theoretische Physik, ETH-H\"onggerberg, CH-8093 Z\"urich, Switzerland}
\address{$^{3}$Department of Electronic Science and Engineering, Kyoto University,
Kyoto 606-8501, Japan}

\date{\today}

\wideabs{

\maketitle

\begin{abstract}
Recent in-plane thermal (Nernst) and interlayer (tunnelling)
transport experiments in Bi$_2$Sr$_2$CaCu$_2$O$_{8+y}$ high
temperature superconductors report hugely different limiting
magnetic fields. Based on pairing (and the uncertainty principle)
combined with the definitions of the Zeeman energy and the
magnetic length, we show that in the underdoped regime both fields
convert to the same (normal state) pseudogap energy scale $T^*$
upon transformation as orbital and spin (Zeeman) critical fields,
respectively. We reconcile these seemingly disparate findings
invoking separated spin and charge degrees of freedom residing in
different regions of a truncated Fermi surface.
\end{abstract}

}

\narrowtext

Establishing the fundamental length and energy scales associated
with the superconducting state in the cuprates is pivotal to
understanding the origin of the high transition temperature $T_c$.
The values of the upper critical field $H_{c2}$ are of particular
importance, for they mark the onset of superconducting correlations and
directly inform on microscopic parameters such as the coherence
lengths in the superconducting state. Early attempts
\cite{Welp,Gray} have demonstrated the difficulty in mapping the
$H_{c2}$ boundary: it lies in part in the high field range
required, but more fundamentally in the large thermal fluctuation
regime \cite{Blatter} and the loss of long-range phase coherence
below $H_{c2}(T)$, making this limiting field `fuzzy' and hard to
pinpoint using the usual experimental tools, e.g., transport or
magnetization. Complicating matters further is the normal state
pseudogap \cite{Millis} which dominates the phase diagram
\cite{Shibauchi} and whose still unresolved connection to
superconductivity is central to the issue of the onset of pairing
and coherence. In this context, the stark difference in the doping
dependencies of the pseudogap energy scale $T^\star$ and $T_c$ has
been well established: $T^\star$ is decreasing (linearly) with
charge doping, while $T_c$ follows the well known superconducting
`dome' described by the phenomenological formula \cite{Bernhard}
$T_c/T_c^{\rm max} = 1-82.6(p-0.16)^2$.

In a view where pairing correlations onset at $T^\star$ and then
acquire global coherence at a lower energy scale $T_c$, the region
$T_c \leq T \leq T^\star$ is a vast fluctuation regime. The
question remains about the doping dependencies of the relevant
magnetic field scales, the field $H_{c2}$ limiting the regime of
superconducting response and the pseudogap closing field $H_{pg}$.
Here we demonstrate the \textit{interconnectedness} of three
magnetic scales: (i) the Zeeman field corresponding to the onset
of spin correlations at the pseudogap scale $T^\star$ and scaling
linear with $T^\star$; this field is identified with the
experimentally observed \cite{Shibauchi} pseudogap closing field
$H_{pg}$, (ii) the orbital critical field $H_{c2}^\star$ quadratic
in $T^\star$, corresponding to the onset of charge correlations
and experimentally determined ($\rightarrow H_{c2}^N$) via thermal
(Nernst) transport measurements \cite{Ong2Hc2}, and (iii) the
upper critical field $H_{c2}$ that marks the onset of global
superconducting coherence and which has been experimentally
tracked ($\rightarrow H_{sc}$) through the presence of large
interplane Josephson currents \cite{Morozov}. This last field,
$H_{sc} \sim H_{c2}$, follows the superconducting `dome' according
to $H_{sc}(p) \sim 1.4 T_c (p)$ and coincides with the usual
unique upper critical field $H_{c2}$ on the strongly overdoped
side of the dome. As we will present below, the distinctly
different orbital ($H_{c2}^\star \sim H_{c2}^N$) and the Zeeman
($H_{pg}$) limiting fields can coexist owing to charge and spin
degrees of freedom separated to different parts of the cuprates'
strongly anisotropic Fermi surface.

Before discussing their interrelation, we first briefly
recapitulate the origin of the three magnetic field scales
$H_{sc}$, $H_{c2}^N$, and $H_{pg}$. In cuprates, the conventional
derivation of the coherence length $\xi$ through an evaluation of
$H_{c2} = \Phi_0 /2\pi \xi^2$ from transport measurements
\cite{Mackenzie,Ando} ($\Phi_0 = hc/2e$ denotes the flux quantum)
has proved notoriously unreliable: a consequence of the presence
of a large vortex liquid regime \cite{Blatter} and the lack of
sharp features in the resistivity. A feature that \textit{can} be
accurately mapped from the field dependence of the interlayer
$c$-axis resistivity $\rho_c(H) = \sigma_c^{-1}(H)$ is the peak at
$H_{sc}$ that corresponds to a crossover from the mostly Josephson
(Cooper pair) tunnelling conductivity $\sigma_J(H)$ at low fields
to that of quasiparticles, $\sigma_q(H)$, at fields above $H_{sc}$
\cite{Morozov}. From the measurements of $\rho_c(H)$ in
Bi$_2$Sr$_2$CaCu$_2$O$_{8+y}$ \cite{Shibauchi,Morozov,Qmelt} we
have found that for \textit{all} doping levels $H_{sc}(T)$ is
nearly $T$-exponential (see inset in Fig.~1).

To understand the temperature dependence of $H_{sc}$ and its
connection to the coherence length $\xi$ we explicitly write the
experimentally established form for the $c$-axis conductivity at
high fields \cite{Morozov}
\begin{equation}
\sigma_c \simeq  \underbrace{\sigma_{J0}
\textrm{exp}~{\left[{{U(H)} \over {T}}\right]}}_{\sigma_J} +
\underbrace{\sigma_{q0}\left(1+ {{H} \over
{H_{c2}^\star}}\right)}_{\sigma_q};
\end{equation}
the two-channel tunnelling process comprises a term $\sigma_J$
controlled by the thermally activated diffusive drift of pancake
vortices hopping over the energy barriers $U(H)$ in the CuO$_2$
planes \cite{Koshelev} and a term $\sigma_q$ due to (nodal)
quasiparticle tunnelling \cite{Morozov}. Here, $\sigma_{J0}$ and
$\sigma_{q0}$ are the corresponding $T\rightarrow 0$ extrapolations
and $H_{c2}^\star \sim \Phi_0 {T^\star}^2/\hbar^2 v_F^2$, where
$T^\star$ coincides with the gap in the quasiparticle spectrum
\cite{Morozov}.
Taking the derivative of Eq.~(1) with respect to $H$ and recalling
that $U(H) \sim U_0 \,\textrm{ln} ({a_0/\xi}) \propto \textrm{ln}
({{H_{c2}}/{H}})$ for a 2D vortex lattice \cite{Feigelman,Triscone},
we obtain the expression for the field $H_{sc}$ where the maximum
in $\rho_c$ (minimum in $\sigma_c$) will occur,
\begin{equation}
H_{sc}(T) \approx H_{c2}\left[{{\sigma_{q0}} \over {\sigma_{J0}}}
{{T} \over {U_{0}}} {{H_{c2}} \over {H_{c2}^\star}}\right]^{-{{T}
/ {U_0}}}.
\end{equation}
Indeed, at low temperatures this high-field $H_{sc}(T)$ can hardly
be distinguished in the experiments from a $T$-exponential
behavior \cite{Koshelev2} [for $x \rightarrow 0$, $x^{-x} =
\textrm{exp}~ {(-x \textrm{ln} x)}\rightarrow 1$] displayed in the
inset of Fig.~1. Hence, in the zero temperature limit
$H_{sc}|_{T\rightarrow 0} \rightarrow H_{c2}(0)$. Experimentally,
we find the zero temperature values of $H_{sc}$ as a function of
hole doping to follow a parabolic dependence proportional to
$T_c(p)$, defining the `coherence dome' shown in Fig.~1.
Consistent with this, a corresponding `dome'-region in the
field-doping ($H-p$) phase space has been recently mapped
\cite{Wen} from the systematic doping dependence of the coherence
length (the size of the vortex core) using detailed magnetization
measurements in La$_{2-x}$Sr$_x$CuO$_4$.

Let us now consider the field scales corresponding to the
pseudogap energy $T^\star$. Recently, Wang \textit{et al.}
\onlinecite{Ong2Hc2} deduced values of an orbital limiting `upper
critical field' $H_{c2}^N$, past which the charge pairing
amplitude should vanish. The $H_{c2}^N$ was extracted
\cite{Ong1Hc2} from a linear extrapolation (to zero) \cite{Maki}
of a remarkably large Nernst signal (attributed to vortex-like
excitations surviving beyond the point where superconducting phase
coherence has been established \cite{Blatter}) and from scaling
arguments \cite{Ong2Hc2}. $H_{c2}^N$ was found to decrease steeply
with increased doping, implying that the Cooper pairing potential
and the superfluid density follow opposite trends versus charge
doping. This led Wang \textit{et al.}\ to a radical interpretation
\cite{Ong2Hc2} of the role of phase fluctuations in the low doping
region. Central to understanding this observation is how the
Nernst-derived $H_{c2}^N$ relates to the gap $T^\star$ observed by
angle-resolved photoemission (ARPES) \cite{Norman,Valla,ZXShen},
as well as by the intrinsic tunnelling \cite{Suzuki} spectroscopies:
pairing correlations are quenched through localization in a magnetic
field $H$ once the magnetic length $a_0=\sqrt{\Phi_0/H}$
drops below the pair correlation length $\xi^\star = \hbar
v_F/\alpha T^\star$. Here, $v_F$ is the Fermi velocity and
$\alpha$ is a numerical of order unity \cite{Tinkham}. Indeed, the
Nernst-derived magnetic field appears to well match this condition
\cite{Ong2Hc2} and hence qualifies as an orbital limiting (or
critical) magnetic field $H_{c2}^N \sim H_{c2}^\star$; as such it
scales {\it quadratically} in $T^\ast$, $\mu_B H_{c2}^N \sim
{T^\star}^2/m v_F^2$, where $\mu_B$ is the Bohr magneton. Note
that within the frame of BCS theory $H_{c2}^\star$ corresponds to
the conventional $H_{c2} = \Phi_0/2\pi\xi^2 \sim \Phi_0
{T_c}^2/\hbar^2 v_F^2$. In the cuprates, however, $T^\star$ and
$T_c$ are taken to define separate length ($\xi^\star$ and $\xi$)
and field scales ($H_{c2}^\star$ and $H_{c2}$) associated with the
appearance of local charge correlations and global phase
coherence, respectively (see \cite{leewen_97} for a discussion of
$H_{c2}$ in underdoped cuprates; the appearance of two
length/field scales has also been discussed within a BCS-Bose
Einstein crossover scenario \cite{levin_02}).

Remarkably, an even higher critical magnetic field $H_{pg} >
H_{c2}^N$ has been derived from $c$-axis interlayer tunneling
transport $\rho_c$ \cite{Shibauchi}. In these experiments, the
recovery of the normal (ungapped) state $c$-axis conductivity
indicates that the pseudogap $T^\star$ closes at a much larger
field scale $H_{pg}$ (nearly twice at low doping, see Fig.~2).
Again, this limiting field relates to the pseudogap energy scale
$T^\star$, this time, however, via the {\it linear} Zeeman
relation $k_B T^\star = g \mu_B H_{pg}$, where $g\sim 2$
\cite{Walstedt} denotes the (spectroscopic splitting) Land\'e
$g$-factor of the Cu$^{2+}$ ions. Correspondingly, one argues that
spin-singlets are unpaired at the pseudogap closing field
$H_{pg}$. Indeed, the field $H_{pg}$ relates to $T^\star$ via the
linear Zeeman scaling irrespective of whether the applied field is
across CuO$_2$ planes or in-plane \cite{Krusin}. The observed
field anisotropy \cite{Yan} is only that of the $g$-factor
\cite{Watanabe}, strengthening the view that the pseudogap is of
spin-singlet origin.

Given the equivalence of the limiting fields $H_{pg}$ and
$H_{c2}^N$ to the same pseudogap energy scale $T^\star$ but via
different routes, `orbital' for $H_{c2}^N$ and `Zeeman' for
$H_{pg}$, we can simply derive how the two fields relate (as a
function of doping $p$),
\begin{equation}
   {{H_{c2}^N}(p)} = {H_{c2}^\star}(p)
    \equiv \alpha^2
    {{\mu_B H_{pg}(p)}\over {m v_F^2}} H_{pg}(p).
    \label{Hc2_Hpg}
\end{equation}
Note that Eq.~(\ref{Hc2_Hpg}) rests only on pairing (and the
uncertainty principle) combined with the definitions of the Zeeman
energy and the magnetic length.

With the Fermi velocity $v_F$ insensitive to doping \cite{ZXShen},
Eq.\ (\ref{Hc2_Hpg}) predicts a simple quadratic relation
$H_{c2}^N(p) \propto H_{pg}^2(p)$. A comparison of $H_{c2}^N
(T\cong 0)$ (see Fig.~2) and $H_{c2}^\star(0)$, by using most
recently measured values for the Fermi velocity \cite{FermiVel}
$v_F \simeq 2$~eV\AA~ and choosing $\alpha \approx 0.6$, yields a
proper collapse of the data in the low doping (underdoped) regime
$p < 0.16$. $H_{c2}^N (0)$ was obtained in two ways. One, by a
simple matching of $H_{c2}^N$ to $H_{sc}(0)$ at $p \gtrsim 0.2$,
where $H_{sc}(0)$ coincides with the usual $H_{c2}(0)$, see Refs.
\protect\onlinecite{Shibauchi,Qmelt}. Another, from the
`universal' $H_{c2}^N(T)/H_{c2}^N(0)$ \textit{vs} $T/T^N_{ons}$
curve implicit in the data of Refs. \protect\onlinecite{Ong1Hc2}
and \protect\onlinecite{Ong3Hc2}; here $T^N_{ons}$ is the onset
temperature of the Nernst response. Close to optimal doping, the
scaled and the measured orbital fields part their ways:
$H_{c2}^\star$ enters the superconducting `dome' while the
$H_{c2}^N$ follows its edge, pointing to a remarkable distinction
between the low- and the high-doping sides \cite{ZXShen}. It
should be remarked that interlayer tunneling transport is a
consistently robust probe of the pseudogap in the underdoped as
well as in the overdoped regimes \cite{Shibauchi}, since it
measures electron tunnelling that is sensitive to the spin
correlations \cite{Krusin}.

Having the two critical fields $H_{c2}^N$ and $H_{pg}$ related to
a single energy scale $T^*$, the question arises how one could
dispose of the same correlation energy twice: via the orbital
route at $H_{c2}^N$ and then again via the Zeeman effect at
$H_{pg} \gg H_{c2}^N$. This `double jeopardy' is naturally
resolved by a strongly anisotropic (truncated) Fermi surface
\cite{ZXShen}, hosting separated charge and spin degrees of
freedom. A generic starting point is the quantum spin-singlet
liquid forming at the energy scale $T^\star$ --- this spin-liquid
groundstate is void of any long range order and competes with the
antiferromagnet \cite{Anderson,Lee,Rice}. Upon doping, the
spin-liquid becomes energetically favorable, charge and spin
degrees of freedom separate and holes are expected to condense on
the spin-liquid background, turning phase coherent at a lower
energy $T_c$. Recently, `cheap' vortices with staggered-flux cores
\cite{Lee} have been argued to destroy coherence above $T_c$ and
qualitatively explain the Nernst data of Ref.\
\onlinecite{Ong2Hc2}, see also \cite{Renner}.

A common feature of these theories is the breakup of the Fermi
surface (FS) into regions describing spin-singlet pairs and
charged holes: the spin-pairing opens up gaps near the $(0,\pi)$
points (the FS corners), cf. Fig.~3 --- the corresponding
pseudogap energy $T^\star$ establishes correlations on the scale
$\xi^\star \sim \hbar v_F/T^\star$. Upon doping, a truncated Fermi
surface appears around the $(\pi,\pi)$ diagonals. When charges
pair up, they draw correlations from the spin-singlet background,
hence spin-singlet pairing at the FS corners and hole-pairing at
the diagonals derive from the same energy scale $T^\star$
\cite{Rice,Honerkamp}, see also \onlinecite{Ioffe}. While the
pairing energy need not necessarily match the underlying energy
scale $T^\star$ of the spin liquid, experiments using scanning
tunnelling microscopy (STM) of the vortex cores \cite{Renner} do
show that this is indeed the case in the underdoped cuprates. We
emphasize that in the described scenario at sufficiently large
doping the separation is ill defined once the spin- and charge
degrees of freedom merge into Fermi-liquid type quasiparticles.

The above considerations naturally lead to two field scales
$H_{pg}$ and $H_{c2}^N$: the charge degrees are connected to the
orbital field $H_{c2}^N$ obtained from the Nernst transport in
Ref.\ \onlinecite{Ong2Hc2}. The \textit{in-plane} Nernst transport
reflects the dissipation due to nodal quasiparticles \cite{Ioffe2}
in the vortex cores, with momenta nearly parallel to ($\pi,\pi$).
Consequently, $H_{c2}^N$ inhibits hole-pairing at the FS
diagonals, but does not destroy the spin-singlet pairs around the
FS corners -- these spin-singlets are unpaired at the much higher
Zeeman field $H_{pg}$. The breakup of the spin-singlets leaves its
trace in the $c$-axis tunnelling experiment \cite{Takenaka}.
Hence, the identification of two limiting magnetic fields
$H_{c2}^N$ and $H_{pg}$ deriving from the same pseudogap energy
scale $T^\star$ via an orbital and a Zeeman relation,
respectively, finds a natural interpretation in terms of a
reconstructed Fermi surface with separated charge and spin degrees
of freedom.

\vspace{1mm}

We thank Vadim Geshkenbein, Alex Koshelev, Maurice Rice, and
Manfred Sigrist for enlightening discussions. Measurements were
performed at NHMFL supported by the NSF Cooperative Agreement
No. DMR-9527035. T.S.\ is supported by a Grant-in-Aid for
Scientific Research from MEXT.


 \normalsize
\begin{figure}[tb]
\begin{center}
\includegraphics[width=\columnwidth]{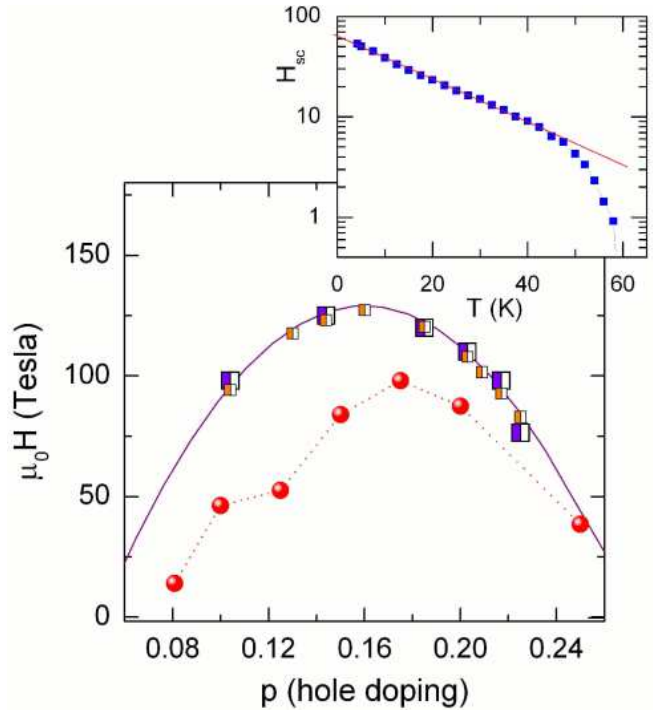}
\end{center}
\vspace{4mm} \caption{(Color online) Doping dependence of the peak
field $H_{sc}$ (half-purple squares) in
Bi$_2$Sr$_2$CaCu$_2$O$_{8+y}$ in the $T \rightarrow 0$ limit.
$H_{sc}(0,p)$ (in Tesla) $\sim 1.4 T_c$ (in Kelvin, shown as
half-orange squares) shapes the `superconducting dome' defined
through the presence of large interplane Josephson currents and
hosting the `conventional' superconducting phase with recombined
quasiparticles. A similar dome-shape is derived from the
magnetization measurements of systematically doped
La$_{2-x}$Sr$_x$CuO$_4$, shown as red dots (from
Ref.~\protect\onlinecite{Wen}). Inset illustrates the nearly
$T$-exponential temperature dependence of $H_{sc}(T)$ in
Bi$_2$Sr$_2$CaCu$_2$O$_{8+y}$ with $p = 0.225$, pointing to the
$T=0$ value of $H_{sc}$. See also Ref.
\protect\onlinecite{Shibauchi}. } \label{fig:1}
\end{figure}

\begin{figure}[tb]
\begin{center}
\includegraphics[scale=0.60]{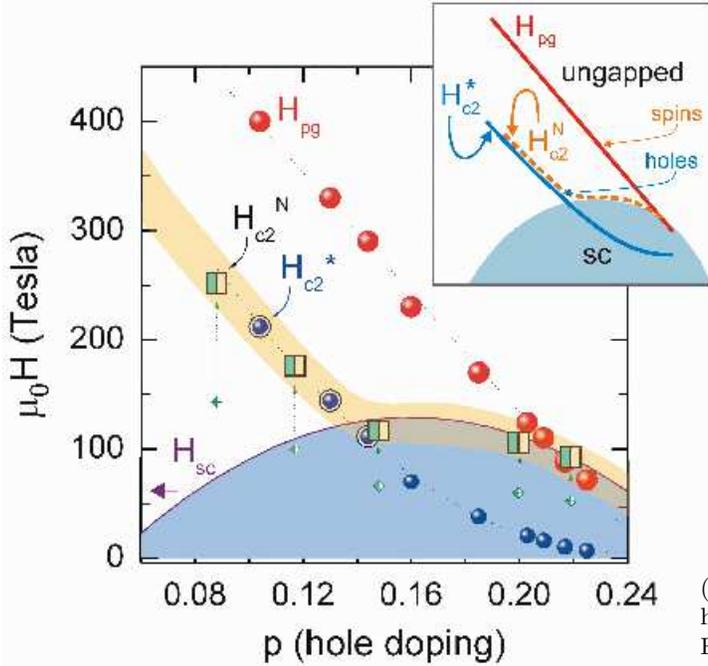}
\end{center}
\vspace{5mm} \caption{(Color online) Comparison of $H_{pg}$,
$H_{c2}^N$ in Bi$_2$Sr$_2$CaCu$_2$O$_{8+y}$, and the
transformation from $H_{pg}$ to $H_{c2}^\star$ via Eq.~(3), where
we used \protect\cite{FermiVel} $v_F \approx 1.8\pm 0.2$~eV\AA~
and $\alpha \approx 0.6$; note the single `$H_{c2}$' scale in the
low doping regime. As sketched for emphasis in the inset, the
lines $H_{c2}^N(0)$ and $H_{c2}^\star(0)$ split apart upon
reaching the superconducting dome (shaded sky-blue) $H_{sc}(0,p)$.
$H_{c2}^N$ data (half-green diamonds) is from scaling near $T_c$
(Ref. \protect\onlinecite{Ong2Hc2}). Light yellow band follows
$H_{c2}^N(0)$ (half-green squares) obtained as described in the
text. } \label{fig:2}
\end{figure}

\begin{figure}[tb]
\begin{center}
\hspace{15mm}
\includegraphics[scale=0.6]{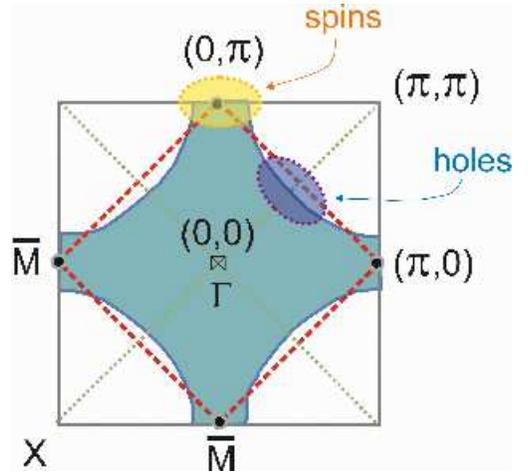}
\end{center}
\vspace{1mm} \caption{(Color online) The breakup of the Fermi
surface (FS) into regions describing spin-singlet pairs and
charged holes: the spin-pairing opens up gaps at the $(0,\pi)$
points (the FS corners) --- the corresponding pseudogap energy
$T^\star$ establishes correlations on the scale $\xi^\star \sim
\hbar v_F/T^\star$. Upon doping, a truncated FS appears around the
$(\pi,\pi)$ diagonals. When charges pair up, they draw
correlations from the spin-singlet background, hence spin-singlet
pairing at the FS corners and hole-pairing at the diagonals derive
from the same energy scale $T^\star$. The separation of spins and
holes will eventually become `fuzzy' in the overdoped regime and
there may be an overlap on the Fermi surface. In this limit the
two (in-plane and out-of-plane) experimental transport probes will
not detect the differences.} \label{fig:3}
\end{figure}

\end{document}